\documentstyle[amssymb,aps,epsf,multicol]{revtex}
\begin{document}
\draft
\author{D. Jonathan, M.B. Plenio and P.L. Knight}
\address{Blackett Laboratory, Imperial College, London SW7 2BZ, United Kingdom}
\title{Fast quantum gates for cold trapped ions.}
\date{\today}
\maketitle

\begin{abstract}
We present an alternative scheme for the generation of a $2$-qubit
quantum gate interaction between laser-cooled trapped ions. The
scheme is based on the AC Stark shift (lightshift) induced by
laser light resonant with the ionic transition frequency. At {\it
specific} laser intensities, the shift of the ionic levels allows
the resonant excitation of transitions involving the exchange of
motional quanta. We compare the performance of this scheme with
respect to that of related ion-trap proposals and find that, for
an experimental realisation using travelling-wave radiation and
working in the Lamb-Dicke regime, an improvement of over an order
of magnitude in the gate switching rate is possible.
\end{abstract}

\begin{multicols}{2}

\section{ Introduction}

The last few years have seen impressive progress in the experimental
demonstration of quantum information processing \cite{JMO}. Among the
growing number of possible physical scenarios for these demonstrations,
the system of laser-cooled trapped ions still remains one of the
most experimentally attractive \cite{Monroe,Monroe2,Turchette,Naegerl} (for
reviews of ion-trap quantum computing, see e.g. \cite{Wineland,Steane,James}%
). Ever since the original ion-trap proposal of Cirac and Zoller (CZ)
 \cite{CZ}, a number of modifications and extensions to their idea have been
proposed
\cite{Turchette,Monroe3,Poyatos,Sorensen,Schneider,Leibfried,Solano}.
Many of these have aimed at bypassing two experimental hurdles of
CZ's proposal, namely: a) cooling the ionic motion to the ground
state, while b) at the same time keeping the ions sufficiently far
apart that individual laser access to each of them is possible. On
the one hand, `hot' gate implementations have been suggested
\cite{Poyatos,Sorensen,Schneider} that aim to function even in the
presence of moderate motional heating. On the other hand, an
ingenious method has been suggested that exploits the ionic
micromotion induced by dc offset potentials to address individual
ions even while simultaneously illuminating all ions with the same
beam
 \cite{Turchette,Leibfried}. Each of these proposals has its own merits and
difficulties, and their feasibility and/or scalability have yet to be
demonstrated in experiment. In the meantime, at least one experiment
currently under development \cite{Naegerl} aims to tackle the two problems
directly, achieving the conditions required by CZ.

In the present paper we assume that these conditions will indeed
become feasible, and focus instead on another aspect of these
experiments: the gate switching rates. Evidently, it is desirable
that these should be as large as possible, so that a reasonably
complex sequence of quantum operations can be realised before
decoherence sets in. It has been remarked \cite{Wineland,Steane},
that the speed of any 2-qubit gate realised by coupling two ions
via a motional mode must be bounded from above by the frequency of
that mode (roughly speaking, the ions must be able to `realise
that they are moving' before they can influence each other). At
the moment of writing, no experiment realising a true 2-qubit
ion-ion gate has been reported. However, at least two experiments
have used schemes similar to the CZ proposal to implement 2-qubit
gates between a single ion and a motional mode \cite{Monroe,Roos}.
Strikingly, in both cases the reported gate speeds fell far short
of the mode frequency, by two to three orders of magnitude. This
limitation was not circumstantial, but inherent in the
experimental technique that was used. The problem was the
existence of strong off-resonant ion-mode transitions, whose
unwanted driving would spoil the desired gate dynamics
\cite{Wineland,Steane,CZ}. In order to avoid this, the laser power
had to be kept at a relatively modest level, resulting in slow
gates. (Very recently, a modification of the CZ scheme which
allows somewhat faster gates has been proposed \cite{Steane00},
see {\it endnote}).

In this paper, we propose a new scheme for 2-qubit gates which
should allow an increase in gate speed by at least an order of
magnitude with respect to these experiments. Furthermore, this
gain is achieved without significant changes in experimental
requirements with respect to existing setups, apart from an
increase in laser power and good intensity stability. The key
feature of our scheme is that it exploits the AC Stark-shift
(lightshift) induced by light resonant with the ionic carrier
transition. Using a coordinate transformation suggested by
Moya-Cessa and co-workers \cite{Hector} we demonstrate that,
within the Lamb-Dicke regime, and at specific shift magnitudes
(i.e, laser intensities), the ion-mode dynamics assumes the form
of a Jaynes-Cummings interaction \cite{Shore}. This interaction
can be exploited to generate a 2-qubit gate in a manner analogous
to the CZ proposal.

We then proceed to compare our scheme with other existing
proposals for faster cold-atom gates. For example, already in
\cite{CZ} it has been pointed out that if the travelling-wave
radiation used in current experiments is replaced with a standing
laser field, with the ion located at a node, then a substantial
increase in gate speed would be possible. The elegant `Magic
Lamb-Dicke parameter' (MLDP) method proposed by Monroe {\it et al}
\cite{Monroe3} could also in principle lead to faster gates. We
argue however that our method, or possibly a combination of it
with the MLDP method, is the one most amenable to practical
implementation within the cold-ion scenario.

The paper is organized as follows: in the first section, we
introduce our gate scheme, explaining its basic principle, the
pulse sequences it requires and the ways in which it differs from
existing schemes. We also discuss its scalability to many-atom
arrays. We then provide numerical confirmation of our analysis,
and compare the performance of our scheme with that of the
Cirac-Zoller scheme in both its regimes (using travelling- or
standing-wave radiation). Finally, we present our conclusions.

\section{2-qubit gates based on the AC Stark-shift effect}

An important feature of the Cirac-Zoller gate scheme \cite{CZ} is
that the frequencies of the pulses it uses are chosen to be resonant with
the transitions between the `bare' (uncoupled) ion-mode levels.
This choice reflects a `perturbative' point of view in which these
level spacings are assumed to be unaffected by the coupling
itself, or in other words that the level shifts due to the AC
potential of the coupling field itself can be disregarded. For a
sufficiently strong field, this assumption breaks down and the
normally disregarded off-resonant transitions become important
(see e.g. \cite{Wineland}, sec. 4.4.6). A number of authors have
speculated that it might be possible to design a gate scheme
incorporating these shifts as an integral feature
\cite{Steane,Cirac96}. In this section we construct a concrete
realisation of this idea, implementing 2-bit gates by exploiting
the lightshift generated by light resonant with the ionic carrier.

\subsection{One ion interacting with a travelling laser field\label{1ionsec}}

In order to present our underlying idea in its clearest form, we
consider first the relatively simple situation of a single trapped
ion interacting with a travelling-wave field. Also for simplicity,
we assume the relevant ionic levels to be coupled by a direct
(optical) transition. As is well-known, the analysis can be
straightforwardly adapted to the case of a Raman two-photon
transition by a suitable redefinition of parameters
\cite{Wineland,James}. In later sections we demonstrate how the
scheme is scalable to traps containing an $N$-ion chain, allowing
2-qubit gates to be realised between the internal states of any
two of the ions.

In the standard interaction representation, the Hamiltonian for
the one-ion system can be written as \cite{Wineland}
\begin{equation}
H=\hbar \Omega \left[ \sigma _{+}\exp \left( i\eta \left[ ae^{-i\nu
t}+a^{\dag }e^{i\nu t}\right] -i\delta t\right) +h.c.\right] .  \label{IP}
\end{equation}
Here, $\delta =\omega _{l}-\omega _{a}$ is the laser-atom detuning, $\nu $
the trap frequency, $\eta =\sqrt{\frac{\hbar k^{2}}{2m\nu }}$ is the
Lamb-Dicke parameter of the trap and we have already taken into account a
rotating-wave approximation (RWA) that assumes $\delta \ll \omega
_{a}+\omega _{l}$ (the detuning is far smaller than optical frequencies).

\begin{figure}[hbt]
\leavevmode
\epsfxsize=8cm
\epsfbox{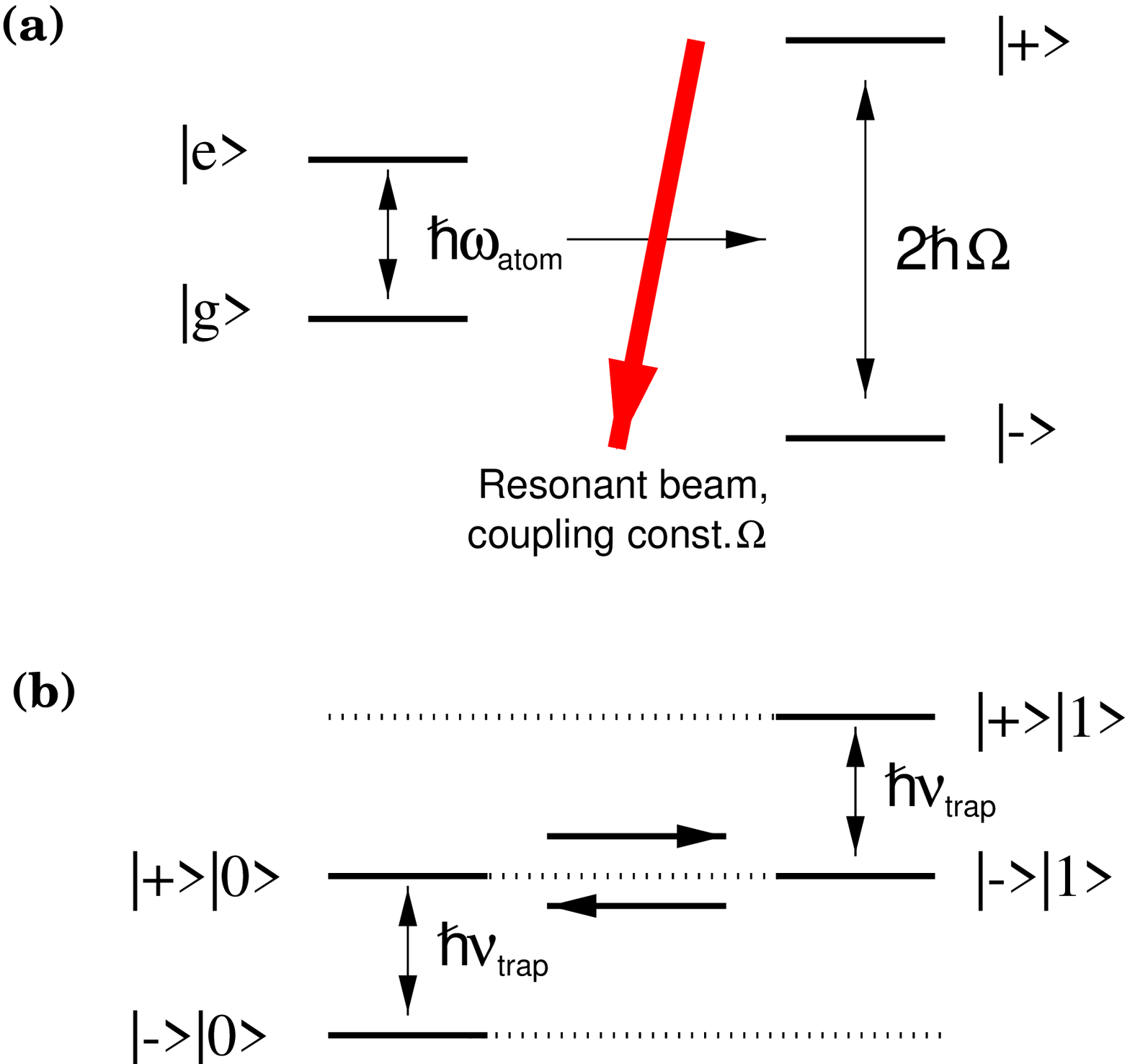}
\caption{\narrowtext Scheme for 2-qubit ion-mode interaction based on the AC
Stark-shift (lightshift) effect. (a) Radiation resonant with the ionic
carrier transition induces a splitting of the dressed levels $|\pm\rangle$
in the interaction picture, by an amount proportional to the laser power.
(b) When the splitting becomes equal to one motional quantum $%
\hbar\nu$, coherent population oscillations are induced between
states $|+\rangle|0\rangle$ and $|-\rangle|1\rangle$.}
\label{fig1}
\end{figure}

Let us briefly recapitulate the approach that is usually taken to
this problem (see, e.g. \cite{Wineland,James,Vogel} and references
therein for
detailed treatments). First, one expands the exponentials in powers of $%
a,a^{\dag }$ and looks for the resonances that arise whenever the
laser frequency is tuned to a motional sideband, i.e., $\delta
=\pm m\nu $. A second RWA is then realised, ignoring off-resonant
terms which rotate at multiples of the trap frequency $\nu$. The
remaining resonant terms can be interpreted in general as
intensity-dependent `multiphonon' transitions \cite{Vogel}.
If the Lamb-Dicke parameter is also small $\left( \eta \ll
1\right) $, and the ion is sufficiently cooled, the intensity
dependence of the coupling constant can be ignored to lowest order
in $\eta $. For example, if the laser is resonant with the carrier
transition ($m=0$), or with the first red sideband ($m=-1$), we
have respectively the simple forms
\begin{mathletters}
\begin{eqnarray}
H_{1CZ} &\simeq &\hbar \Omega e^{-\frac{1}{2}\eta ^{2}}\left[
\sigma _{+}+\sigma _{-}\right]  \label{1bit} \\ H_{2CZ} &\simeq
&i\hbar \Omega \eta e^{-\frac{1}{2}\eta ^{2}}\left[ \sigma
_{+}a-\sigma _{-}a^{\dagger }\right] \; .  \label{2bit}
\end{eqnarray}
These are the interactions that form the basis of the standard
Cirac-Zoller scheme for realising 1- and 2-qubit quantum logic
gates \cite{CZ}. A slight modification of this scheme (using
blue-sideband-detuned pulses) has been implemented experimentally
in single-ion traps \cite{Monroe,Roos}.

\subsection{2-qubit lightshift-based quantum gates}

We now demonstrate that, even if only radiation resonant with the
carrier is used, and without leaving the Lamb-Dicke limit, there
is still a regime where 2-qubit dynamics can be obtained. The
basic physical idea behind this is as follows: we know that, apart
from driving the 1-qubit transition described in Eq.(\ref{1bit}),
any radiation resonant with the carrier will also lead to an AC
level-splitting of the ionic semiclassical dressed states $\left|
\pm \right\rangle =\frac{1}{\sqrt{2}}\left( \left| g\right\rangle
\pm \left| e\right\rangle \right) $ (Fig. \ref{fig1}$a$)
\cite{anytextbook}). The magnitude of the splitting is $2\hbar \Omega $, where $%
\Omega $, is the Rabi frequency. When the intensity of the laser
is such that the splitting equals {\it exactly%
} one vibrational energy quantum $\hbar v$, the levels $\left|
+\right\rangle \left| 0\right\rangle $ and $\left| -\right\rangle
\left| 1\right\rangle $ become degenerate, and we can expect
transitions between them (Fig. \ref{fig1}$b$). This amounts
effectively to an exchange of excitation between the motional and
internal states, i.e., to 2-qubit dynamics.

To see how this happens in detail, let us begin by first making the
Lamb-Dicke approximation (to first order in $\eta $) directly in eq. $\left(
\ref{IP}\right) $%
\end{mathletters}
\begin{eqnarray}
H &\simeq& \hbar \Omega e^{-\frac{1}{2}\eta ^{2}}\left[ \sigma
_{+}e^{-i\delta t}\left( 1+i\eta \left[ a e^{-i\nu t}+a^{\dag } e^{i\nu t} \right]
\right) +h.c.\right]  \nonumber \\
&=&\hbar \Omega^{\prime} \left[
\begin{array}{c}
\left( \sigma _{+}e^{-i\delta t}+\sigma _{-}e^{+i\delta t}\right) + \\
i\eta \left( \sigma _{+}e^{-i\delta t}-\sigma _{-}e^{+i\delta t}\right)
\left[ a e^{-i\nu t}+a^{\dag } e^{i\nu t} \right]
\end{array}
\right]  \label{LD}
\end{eqnarray}
(where we have defined $\Omega ^{\prime }\equiv \Omega e^{-\frac{1}{2}\eta
^{2}}$). When the radiation is resonant with the ionic transition (or `carrier')
frequency $\left( \delta =0\right) $, this reduces to
\begin{equation}
H \simeq \hbar \Omega ^{\prime }\left[ \sigma _{+}+\sigma _{-}
+i\eta \left( \sigma _{+}-\sigma _{-}\right) \left[ a e^{-i\nu
t}+a^{\dag } e^{i\nu t} \right] \right] \; .  \label{1ionres}
\end{equation}
Comparing with eq. (\ref{1bit}), we see that the usual derivation
corresponds to neglecting the terms rotating at frequency $\pm \nu
$ in this expression. These terms are the first order correction
to the semiclassical ion-field interaction due to the presence of
the trapping potential, and their effect is to cause the dressed
states $\left| \pm \right\rangle $ to become nonstationary. To see
how these evolve, we first move into the `dressed-state' picture
obtained by rotating the atomic basis states with the
transformation
\begin{equation}
R=\frac{1}{\sqrt{2}}\left(
\begin{array}{ll}
1 & 1 \\
-1 & 1
\end{array}
\right) ,  \label{R}
\end{equation}
so that $\left| \pm \right\rangle $ become respectively $\left|
e\right\rangle $ and $\left| g\right\rangle $ (note that, in our notation, $%
\left| e\right\rangle =\left(
{1 \atop 0} \right) $, $\left| g\right\rangle =\left(
{0 \atop 1} \right)$, $\sigma_+={0\,1\choose0\,0}$,
\mbox{$ \sigma_-={0\,0\choose1\,0} $}).

Using the fact that
\begin{equation}
R\sigma _{\pm }R^{\dag }=\frac{1}{2}\left( \sigma _{z}\pm \left( \sigma
_{+}-\sigma _{-}\right) \right)
\end{equation}
we can see that, in this picture, the Hamiltonian has the Jaynes-Cummings
\cite{Shore} form
\begin{equation}
H'=\hbar \Omega ^{\prime }\left[ \sigma _{z} +i\eta \left( \sigma _{+}-\sigma
_{-}\right) \left[ a e^{-i\nu t}+a^{\dag } e^{i\nu t} \right] \right] .  \label{LDJCM}
\end{equation}
(This transformation of the Hamiltonian is a special case of the
construction given in \cite{Hector}, where it is shown that the ion-laser
interaction is {\em always} unitarily equivalent to a Jaynes-Cummings form,
without any approximations).

Making a further `interaction picture' transformation of the Hamiltonian by
the unitary operator $\exp \left( \frac{i\Omega ^{\prime }t\sigma _{z}}{%
\hbar }\right) $, we have
\begin{equation}
H''=i\hbar \eta \Omega ^{\prime }\left[
\begin{array}{c}
e^{i\left( 2\Omega ^{\prime }-\nu \right) t}\sigma _{+}a-e^{-i\left( 2\Omega
^{\prime }-\nu \right) t}\sigma _{-}a^{\dag } \\
+e^{i\left( 2\Omega ^{\prime }+\nu \right) t}\sigma _{+}a^{\dag
}-e^{-i\left( 2\Omega ^{\prime }+\nu \right) t}\sigma _{-}a
\end{array}
\right] ,  \label{dressedpic}
\end{equation}
which gives us the resonance condition
\begin{equation}
\Delta =\Omega ^{\prime }-\frac{\nu }{2}=0.  \label{resonance}
\end{equation}
Apart from the small correction to $\Omega $ given by the Debye-Waller
factor $e^{-\frac{1}{2}\eta ^{2}}$ \cite{Wineland}, this is precisely the
condition depicted in Fig. \ref{fig1}. In this case, the first two
(`rotating') terms in Eq. (\ref{dressedpic}) become constant while the
second two (`counter rotating') oscillate at frequency $2\nu $. We can
ignore them, making the Jaynes-Cummings RWA, as long as the secular
frequency $\eta \Omega ^{\prime }=\frac{1}{2}\eta \nu $ of the resulting
evolution is much smaller than this \cite{Shore}. This requires $\eta \ll 4$%
, which is compatible with the Lamb-Dicke assumption $\eta \ll 1$
we have already made. Thus, if the laser's frequency and intensity
are such that they satisfy the {\it %
double resonance} condition $\delta =\Delta =0$, the evolution of the system
can be described by the simple Jaynes-Cummings form
\begin{equation}
H_{2SS}=\frac{i\hbar \eta \nu }{2}\left[ \sigma _{+}a-\sigma _{-}a^{\dag
}\right] .  \label{2bitSS}
\end{equation}

\begin{figure}[hbt]
\leavevmode \epsfxsize=15cm \epsfbox{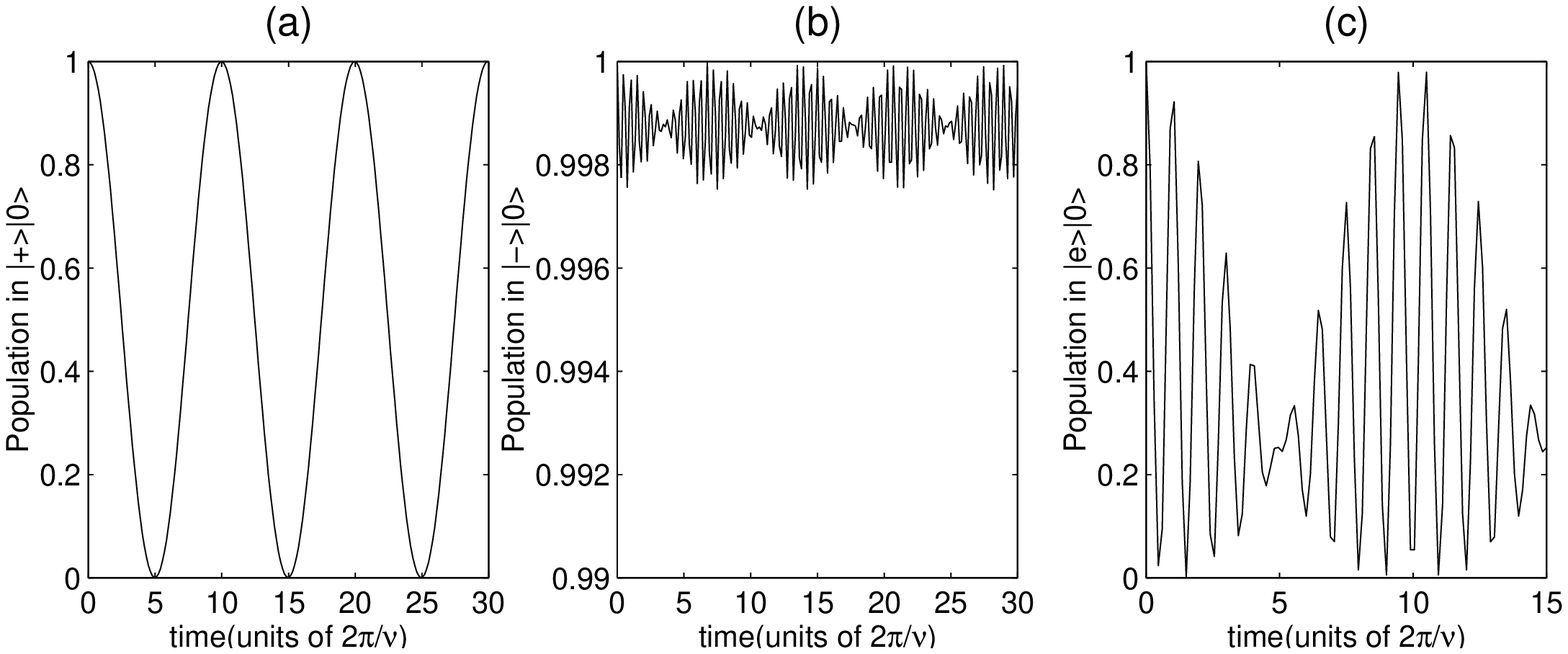} \caption{\narrowtext
Simulation of a lightshift-based ion-mode gate operating on a
single trapped ion. The laser is tuned on resonance with the
carrier, and also at an intensity such that the Rabi frequency
$\Omega^{\prime}$ is equal to exactly half the trap frequency
$\nu$. The Lamb-Dicke parameter is $\eta=0.1$. (a) In the
interaction picture, state $|+\rangle| 0\rangle$ exchanges
population with $|-\rangle| 1\rangle$; an exchange rate of over
99\% is achieved. (b) Meanwhile, state $|-\rangle| 0\rangle$ is
stationary. The resulting ion-mode ``conditional dynamics" can be
used to implement a 2-qubit quantum gate. (c) In the
Schr\"{o}dinger picture, this effect appears as a modulation of
the Rabi oscillations between states $|e\rangle| 0\rangle$ and
$|g\rangle| 0\rangle$.} \label{ACfig}
\end{figure}

What this teaches us is that off-resonant transitions cannot
always be disregarded, but, under the right conditions, may in
fact lead to resonant effects. Intuitively, if the off-resonant
terms in the Hamiltonian given in eq. (\ref{1ionres}) rotate
precisely in step with the secular evolution generated by the
resonant terms, their contribution does not `average out' but
rather adds up over each cycle, in a manner reminiscent of an
oscillator being driven by a resonant force. In the present case
this effect allows a field resonant with the carrier to couple the
internal and motional ionic variables in a way exactly analogous
to a red-sideband-detuned pulse as described by eq. (\ref{2bit})
(Fig. \ref{ACfig}$a,b$). In particular, it can just as well be
used to implement $2$-qubit logic gates between these two degrees
of freedom. Of course, the Hamiltonian (\ref{2bitSS}) is valid
only in the `dressed' picture defined by the operator $R$ in eq.
(\ref{R}). In the normal or `bare' picture, its effect can be seen
as a beating at frequency $\eta \nu $ superposed on the usual Rabi
flops between states $\left| g\right\rangle \left| n\right\rangle
$ and $\left| e\right\rangle \left| n\right\rangle $ (Fig.
\ref{ACfig}$c$). It is not necessarily obvious that this
`dressed-picture' Jaynes-Cummings interaction can be used to
implement quantum logic gates in the `real world'. Nevertheless,
in the next section we show how, with a suitable generalisation to
the $N$-ion situation, this interaction can indeed realise a
Control-NOT (C-NOT) gate between the internal variables of
two separate ions. (Recall that a C-NOT gate together with
one-qubit rotations form a universal set of gates for quantum
computing \cite{Barenco}).

Finally, let us briefly consider the experimental requirements of
our proposal. Apart from the usual demands of the CZ quantum gate
proposal (individual ion access, ground-state cooling), the only
new requirement we make is that the laser should have a fixed
intensity satisfying the resonance condition in eq.
(\ref{resonance}) (or its $N$-ion generalisation, see below). In
more quantitative terms, our numerical simulation (see section
\ref{numressec}) indicates that the laser power must be stable to
within about $\pm 0.5\%$. This does not seem to require
significant improvements in the laser power and intensity
stability already available in current experimental setups
\cite{Christoph}. There is also a bonus in the fact that a single
laser can be used to perform both 1- and 2- qubit interactions.
Therefore, we expect that a proof-of-principle experiment using a
single trapped ion should not be hard to realise.

\subsection{Lightshift gates in a chain of N ions}

\label{Nbitsec}

The results we have just described are almost immediately
generalisable to the case where there are $N$ identical ions (and
therefore $N$ motional modes) in a linear trap
\cite{Wineland,James}. Assuming that each of the ions can be
illuminated individually by a (travelling) laser beam, then
resonance conditions similar to eq. (\ref{resonance}) turn out to
exist for each separate mode frequency $\nu_{j}$. Before showing
how the resulting ion-mode interaction can be used to implement
ion-ion gates, we would like to call attention to an important
aspect of the $N$-ion situation. In principle, any of the $N$
motional modes can be used to couple the internal ionic variables.
However, in order for the lightshift scheme to function with
higher-order modes it is necessary to drive the system deeper into
the Lamb-Dicke regime. To see this, consider the
interaction-picture Hamiltonian describing the coupling of the
$j^{th}$ ion with a (travelling-wave) laser \cite{James}
\begin{equation}
H=\hbar \Omega \left[ \sigma _{+}^{j}\exp \left( i
\sum_{p=1}^{N}\eta_{jp}\left[ a_{p}e^{-i\nu _{p}t}+a_{p}^{\dag }e^{i\nu
_{p}t}\right] -\delta t\right) +h.c.\right]  \label{Nionsfull}
\end{equation}
Here, $p$ indexes the normal modes. The parameter $\eta _{jp},$which
functions as the `effective' Lamb-Dicke parameter of the $p^{th}$ mode$,$
corresponds to the product $\eta _{p}b_{j}^{\left( p\right) }$, where $\eta
_{p}=\sqrt{\frac{\hbar k^{2}}{2m\nu _{p}}}$ is the `conventional' Lamb-Dicke
parameter and $b_{j}^{\left( p\right) }$is the relative weight of the $%
j^{th} $ ion's displacement in this mode. For the centre-of-mass mode, $%
b_{j}^{\left( 1\right)} =\frac{1}{\sqrt{N}}$ is independent of which ion is
being driven. For other modes this is no longer true.
James \cite{James} has given values of $b_{j}^{\left( p\right) }$
for all ions and modes up to $N=10$.

If all modes are suitably cooled and within the Lamb-Dicke regime, and
if the laser is resonant with the ionic carrier transition
 $\left( \delta=0\right) $,
then a procedure entirely analogous to the one described in eqs. (\ref{LD}-%
\ref{dressedpic}) can be followed. One then obtains that, in the
`dressed-state' picture defined by
\begin{equation}
V\left( t\right) =\frac{1}{\sqrt{2}}\exp \left( i\Omega ^{\prime }t\sigma
_{z}^{j}\right) R_{j},  \label{V(t)}
\end{equation}
the Hamiltonian given above can be rewritten as
\begin{equation}
H''=i\hbar \Omega ^{\prime }\sum_{p}\eta _{jp}\left[
\begin{array}{c}
e^{i\left( 2\Omega ^{\prime }-\nu _{p}\right) t}\sigma
_{+}^{j}a_{p}-e^{-i\left( 2\Omega ^{\prime }-\nu _{p}\right) t}\sigma
_{-}^{j}a_{p}^{\dag }+ \\
+e^{i\left( 2\Omega ^{\prime }+\nu _{p}\right) t}\sigma _{+}^{j}a_{p}^{\dag
}-e^{-i\left( 2\Omega ^{\prime }+\nu _{p}\right) t}\sigma _{-}^{j}a_{p}
\end{array}
\right] ,  \label{NionHam}
\end{equation}
where $\Omega ^{\prime }=\Omega e^{\frac{-1}{2}\left( \sum_{p}\eta
_{jp}^{2}\right) }$. As expected, there are multiple resonance conditions
analogous to eq. (\ref{resonance}), one for each mode frequency $\nu _{p}.$
If any of these are met (say, $\Omega ^{\prime }=\frac{\nu _{q}}{2}$ for the
$q^{th}$ mode), then the terms in this Hamiltonian can be divided into three
categories according to their time-dependence: (1) The rotating terms of the
$q^{th}$ mode are resonant, and represent a Jaynes-Cummings interaction of
the form
\begin{equation}
H_{2SS}=\frac{i\hbar \nu _{q}\eta _{jq}}{2}\left( \sigma
_{+}^{j}a_{p}-\sigma _{-}^{j}a_{p}^{\dag }\right) ;  \label{effJCM}
\end{equation}
(2) All counter-rotating terms oscillate at frequencies equal (in modulus)
to at least $\nu _{q}+\nu _{1}\gg \frac{\nu _{q}\eta _{jq}}{2}$, where $\nu
_{1}$ is the lowest energy mode. Assuming the effective Lamb-Dicke parameter
$\eta_{jq}$ is small $(\eta_{jq}\lesssim\frac{1}{10})$, they can therefore
can be discarded in a RWA. (3) The rotating terms of the other modes
oscillate at frequencies equal to $\pm \left| \nu _{p}-\nu _{q}\right|$. For
a similar $\eta _{jq}$ these terms can be discarded as long as
\begin{equation}
\frac{\left| \nu _{p}-\nu _{q}\right| }{\nu _{q}}\gg \frac{\eta_{jq}}{2}\;.
\label{cond}
\end{equation}
If this is true for all $p\neq q$, then the Hamiltonian (\ref{NionHam}) can
be reduced to the resonant term given in eq. (\ref{effJCM}). In this case,
only the $q^{th}$ mode is coupled to the ion's internal state, just as in
the usual perturbative scheme when the laser is tuned to the first red
sideband of this mode. The off-resonant terms will lead to a small
population leakage into the unwanted modes, of order
\begin{equation}
\epsilon^2 = \left(\frac{\eta_{jq}\nu_q}{2\left| \nu _{p}-\nu
_{q}\right|}\right)^2 \ll 1 . \label{epsilon}
\end{equation}

As we discuss in appendix A, for high enough precision (small
enough $\epsilon$), this population loss gives an upper bound to
$\eta_{jq}$, and therefore to the overall Rabi frequency
$\frac{1}{2}\eta_jq \nu_q$ at which the scheme can function. For
example, in the case of the lowest (center-of-mass) mode,
$\epsilon^2 \leq 0.005$ requires $\eta_1 \lesssim 0.1$. In
addition, it has been shown by James \cite{James} that the spacing
$|\nu_{q+1}-\nu_q|$ between successive modes decreases as their
order increases. It follows that attaining a given precision
$\epsilon$ requires $\eta_{jq}$ to be made smaller and smaller as
$q$ grows. In effect, we find that the potential increase in Rabi
frequency afforded by using higher modes is completely
counterbalanced by this requirement, with the result that the
maximum value for the overall switching rate actually decreases as
higher modes are used.

\subsection{2-ion CNOT gates}

Assuming the effective Hamiltonian (\ref{effJCM}) is valid, we can
use it to implement 2-qubit quantum logic gates between two ions
in a manner similar to the usual Cirac-Zoller (CZ) scheme
\cite{CZ}. The analogy is not perfect because in the present case
the Jaynes-Cummings Hamiltonian $H_{2SS}$ is valid only in the
picture defined by the unitary operator in eq. (\ref {V(t)}),
which varies according to which atom is being addressed.
Before we realise a gate, we must first transform back into the
`common' picture (i.e., the one where the Hamiltonian in eq.
(\ref {Nionsfull}) is defined) and see how the time evolution
behaves there. In this case we have that an initial state $\left|
\psi \left( 0\right) \right\rangle $ evolves according to
\begin{equation}
\left| \psi \left( t\right) \right\rangle =V^{\dag }\left( t\right)
U_{JCM}\left( t\right) V\left( 0\right) \left| \psi \left( 0\right)
\right\rangle
\end{equation}
where $V\left( t\right) $ is given in eq. $\left(\ref{V(t)}\right)$ and $%
U_{JCM}\left( t\right) =\exp \left( \frac{-it}{\hbar
}H_{2SS}\right)$. In particular, the following states have a
simple time evolution:
\begin{eqnarray}
\left| -\right\rangle \left| 0\right\rangle &\rightarrow &\exp \left(
{\displaystyle {i\nu _{q}t \over 2}}%
\right) \left| -\right\rangle \left| 0\right\rangle  \label{evol1} \\
\left| +\right\rangle \left| 0\right\rangle &\rightarrow &e^{\frac{-i\nu
_{q}t}{2}}\cos \left( \frac{\nu _{q}\eta _{jq}t}{2}\right) \left|
+\right\rangle \left| 0\right\rangle -  \label{popflip} \\
&&-e^{\frac{i\nu _{q}t}{2}}\sin \left( \frac{\nu _{q}\eta _{jq}t}{2}\right)
\left| -\right\rangle \left| 1\right\rangle  \nonumber \\
\left| -\right\rangle \left| 1\right\rangle &\rightarrow &e^{\frac{i\nu _{q}t%
}{2}}\cos \left( \frac{\nu _{q}\eta _{jq}t}{2}\right) \left| -\right\rangle
\left| 1\right\rangle + \\
&&+e^{\frac{-i\nu _{q}t}{2}}\sin \left( \frac{\nu _{q}\eta _{jq}t}{2}\right)
\left| +\right\rangle \left| 0\right\rangle  \nonumber \\
\left| +\right\rangle \left| 1\right\rangle &\rightarrow &e^{\frac{-i\nu
_{q}t}{2}}\cos \left( \frac{\nu _{q}\eta _{jq}t}{\sqrt{2}}\right) \left|
+\right\rangle \left| 1\right\rangle -  \label{evol4} \\
&&-e^{\frac{i\nu _{q}t}{2}}\sin \left( \frac{\nu _{q}\eta _{jq}t}{\sqrt{2}}%
\right) \left| -\right\rangle \left| 2\right\rangle.  \nonumber
\end{eqnarray}
As we can see, we obtain the usual Jaynes-Cummings Rabi flops, except that
here the atomic states for which the atom and mode dynamically entangle and
disentangle themselves are the dressed states $\left| \pm \right\rangle $,
not the bare states $\left| g\right\rangle ,\left| e\right\rangle $. There
are also some additional time-dependent phases.

In Appendix B, we demonstrate explicitly how this evolution can be used to
implement a 2- qubit gate between two ions. We follow the same basic
three-step pulse sequence proposed by Cirac and Zoller \cite{CZ}: first, a $%
\pi $-pulse is realised between ion 1 and the chosen vibrational
`data bus' mode, which is initially cooled to the ground state.
This effectively maps the internal state onto the motional one and
vice-versa, implementing the so-called SWAP gate. Second, a $2\pi
$-pulse is applied between the mode and ion 2, realising an
entangling gate between the two systems. Finally, a second $\pi
$-pulse maps the motional state back onto the first ion,
completing the ion-ion gate. In the lightshift scheme, some minor
modifications in the sequence are necessary due to the fact that
the `computational basis states' of the ions (generally assumed to
be the bare states $\left| g\right\rangle ,\left| e\right\rangle
$) are not favoured by the time-evolution above. This will then
require a few extra $1$-qubit rotations in between the three basic
steps. In the end, we are able to implement a C-NOT gate, with ion
2 acting as the `control' qubit, using a sequence of six pulses
(three 1-qubit and three 2-qubit pulses). In comparison, the
original CZ proposal requires $5$ pulses to implement a C-NOT
gate, with the ions assuming the opposite roles: ion 1 is the
`control' and ion 2 the `target' qubit. We note that, in our
protocol, some of the 1-qubit pulses may (at least in principle)
be realised simultaneously with a 2-qubit pulse: pulses 1 and 2 in
Appendix B can realised together, and the same is true of pulses 4
and 5. In contrast, in the CZ scheme each of the five pulses must
be realised in sequence.

\section{Comparative performance of gate schemes}

We now study the performance of our `lightshift-based' (LB) gate
scheme, comparing it to that of Cirac and Zoller's original
`red-sideband pulse' proposal \cite{CZ}. Briefly speaking, our
goal is to estimate the overall switching rate for an ion-ion
C-NOT gate that can likely be attained using each scheme.

We begin by recalling that this rate will be essentially governed
by the speed of the three 2-qubit steps in either scheme's pulse
sequence. This follows since 1-qubit ionic gates are unlimited by
the mode frequency, and can therefore be implemented at a much
greater speed than 2-qubit ion-mode pulses \cite{Wineland}. If we
also assume for simplicity that the same ionic transition is used
for both $\pi-$ and $2\pi-$pulses, then the overall ion-ion gate
frequency should be approximately equal to the 2-qubit
Jaynes-Cummings Rabi frequency. Here we are using the convention
that one complete Rabi oscillation, i.e., when all states
{\it and} their phases have returned to their initial values, corresponds
 to a $4\pi$ pulse.

For the LB scheme, this frequency is just $\frac{\eta\nu}{2}$. For
a typical value $\eta=0.1$ of the Lamb-Dicke parameter, we obtain
therefore an overall C-NOT switching rate of about
$\frac{\nu}{20}$. Although still well under the limit posed by the
mode frequency $\nu$ itself, such a rate would represent a
substantial improvement with respect to current experiments. For
example, in the (single-ion) 2-qubit gate experiment reported in
\cite{Monroe}, the 2-qubit Rabi frequency was approximately
$10^{-3}\nu$. In what follows, we elaborate on this comparison by
making a more thorough analysis of the limits of validity of the
two methods. In particular, we include numerical confirmation of
the efficiency of the LB scheme.

\subsection{Regimes of the Cirac-Zoller scheme}
Unlike in the LB scheme, in the CZ method the speed of the 2-qubit
gates is directly proportional to the laser field used to drive
the red-sideband transition. This field cannot however be made too
intense without driving unwanted off-resonant transitions, which
therefore are the limiting factor on the resulting gate speed.
Before we can properly assess this limit quantitatively, we must
first recall that the CZ scheme operates in two strikingly
different regimes, depending on the spatial profile of the laser
field \cite{CZ,Cirac96}. The origin of this difference lies in the
presence or not of strongly coupled off-resonant levels. It turns
out that the conditions under which transitions to these levels
can be safely ignored (as is implied in the derivation of the CZ
scheme) depend crucially whether {\it travelling-wave} or {\it
standing-wave} laser radiation is employed to drive the
red-sideband transition

\begin{figure}[hbt]
\leavevmode
\epsfxsize=6cm
\epsfbox{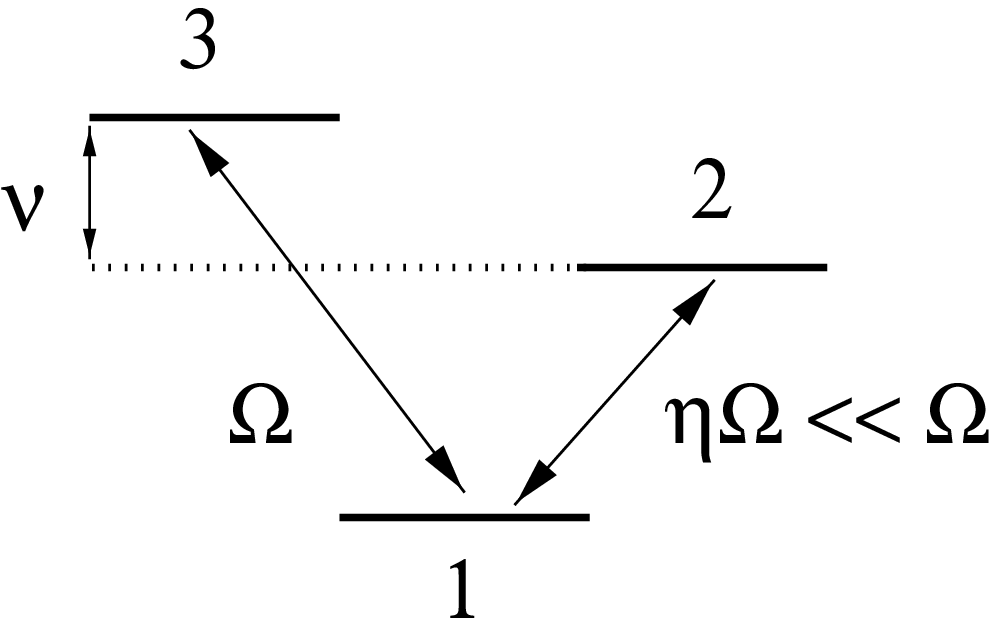}
\caption{\narrowtext Driving an ion with travelling radiation
detuned to the first red sideband generates a situation in many
ways analogous to a 3-level system. The resonant $1
\leftrightarrow 2$ transition corresponds to the relatively weak
sideband transition (coupling constant $\eta\Omega$), while the
off-resonant $1 \leftrightarrow 3$ transition is analogous to the
strong carrier transition (coupling constant
$\Omega\gg\eta\Omega$). The off-resonant transition can be
ignored, leaving an effective 2-level system formed by levels 1
and 2, only if $\Omega$ satisfies the condition in eq. (\ref{travcond}).}
\label{toyfig}
\end{figure}

When a travelling beam is used, the closest-lying off-resonant
transition is the carrier transition itself, which is detuned by
the mode frequency $\nu $. Despite this, the carrier is also {\it
stronger} than the resonant transition by a factor of
$\eta^{-1}\gg 1$. Intuitively, this situation is analogous to a
V-type 3-level atom where a weak transition (of strength
$\eta\Omega'$) is being resonantly driven, and where there is
another closely lying transition, detuned by $\nu$, which has a
much stronger coupling constant $\Omega'$ (Fig \ref{toyfig}). The
effects of both transitions must then be carefully weighed against
each other: if $\nu $ is large with respect to $\eta\Omega'$, then
we may expect the off-resonant transition to be `washed out' on
average, as happens in usual rotating-wave approximations.
However, this condition alone is not sufficient, since in the
limit $\eta \rightarrow 0$ the off-resonant transition must
dominate the time evolution, resulting in oscillations with
effective Rabi frequency $\frac{\Omega'^2}{\nu}$. We can therefore
expect the resonant transition to dominate only if its secular
Rabi frequency $\eta\Omega'$ is much greater than this value, i.e.
if
\begin{eqnarray}
\Omega'\ll\eta\nu .
\label{travcond}
\end{eqnarray}
The validity of this heuristic argument for the actual
Cirac-Zoller Hamiltonian can be confirmed via a straightforward
perturbation-theory calculation \cite{Note}.

In other words, in order to ignore off-resonant transitions the
Rabi frequency $\Omega'$ of the ion-mode interaction must be
extremely small, of the order $\nu/100$ for a typical value
$\eta=0.1$. This in turn implies that the switching rate of the
resulting logic gates will be of order $\eta\Omega'\lesssim
\nu/1000$, way below the upper limit set by $\nu$. It is
worthwhile to note that eq. (\ref{travcond}) was indeed satisfied
in both published experiments that implemented CZ-like Rabi flops
using travelling-wave radiation and a single trapped ion
\cite{Monroe,Roos}.

A very different situation arises if the laser field forms a
sinusoidal standing wave (such as could be obtained by bouncing the
beam back on itself from a mirror), and if the ion is located exactly in one
of the {\it nodes} of this wave. In this case, interference from the two
travelling components of the wave completely cancels many of the
off-resonant transitions, in particular the carrier \cite{Wineland,James,CZ}.
This effective selection rule greatly increases the laser power that can be used,
since the most important off-resonant terms remaining in the Hamiltonian
(Jaynes-Cummings counter-rotating terms and terms describing the
accidental driving of the wrong modes) are
no longer stronger than the resonant one. Standard perturbation-theoretic arguments
\cite{Cirac96,James} show that in this case the laser power should satisfy
\begin{eqnarray}
\Omega'\ll\frac{\nu}{\eta}.
\label{standcond}
\end{eqnarray}
For $\eta=0.1$, this implies an increase by two orders of magnitude
with respect to the travelling-wave case.
As a result, this configuration could potentially lend itself to
the implementation of much faster gates than the ones already
achieved experimentally. Unfortunately, the technical difficulty
of reliably maintaining an ion precisely in a wave node seems to
have discouraged researchers from attempting such an experiment
\cite{Berkeland}. We are also not aware of any current plans for
experiments in this direction.

\subsection{Efficiency of gate implementations}
\label{efficsec} In what follows, we compare our `lightshift-based'
proposal to both regimes of the CZ scheme. We find that
its performance can approach that of the standing-wave CZ
configuration, without the latter's technical drawbacks. In other
words, an improvement of over an order of magnitude in the
switching rate can be achieved with respect to current travelling-wave-based
experiments without a great change in the experimental setup itself. It must
be emphasised again that we are only interested here in the {\it
theoretical} limits to the gate performance, arising exclusively
from the existence of stray off-resonant excitations in the
system. In other words, we are not concerned with external noise
or dissipative effects such as spontaneous emission \cite{Plenio},
but with the maximum performance obtainable even under ideal
experimental conditions.

A useful figure of merit for comparing the performance of the
different schemes can be defined as follows. First, we determine
how efficiently the $\pi$-pulse (or `SWAP gate') step is
implemented in each scheme as a function of a relevant external
parameter of the system, for instance laser power (a precise
definition of what me mean by `efficiency' is given below). We can
then define the {\it maximum switching rate} for each scheme as
the greatest speed that can be attained while simultaneously
keeping the efficiency above a sufficiently high threshold, which
we (arbitrarily) set at 99\%.

The definition of `efficiency' is also somewhat arbitrary. We take
it to be the {\it average fidelity} with which the SWAP gate
operates, maximized over one cycle, or
\begin{equation}
F(\eta,\Omega) = \left.\max\right|_{\mbox{1st cycle}}
\frac{1}{n}\sum_{k=1}^n\left|\langle\psi^k_f| U(\eta,
\Omega,t)|\psi^k_i\rangle\right|^2,
\end{equation}
where the average is taken over some set of `relevant' initial
states $\{\psi^k_i\}_{k=1}^n$, with ideal images under SWAP given by
$\{\psi^k_f\}_{k=1}^n$, and where $U(\eta, \Omega,t)$ represents
the full time evolution of the ion-trap system. For simplicity,
we take this set to be the basis states $\left\{ \left| g\right\rangle
\left| 0\right\rangle, \left| g\right\rangle \left|
1\right\rangle\right\}$ (in the case of the CZ gate) or
$\left\{\left| -\right\rangle \left| 0\right\rangle, \left|
-\right\rangle \left| 1\right\rangle \right\}$ (in the case of the
LB gate).

\subsubsection{Numerical results}\label{numressec}
\begin{figure}[hbt]
\leavevmode \epsfxsize=12cm
\epsfbox{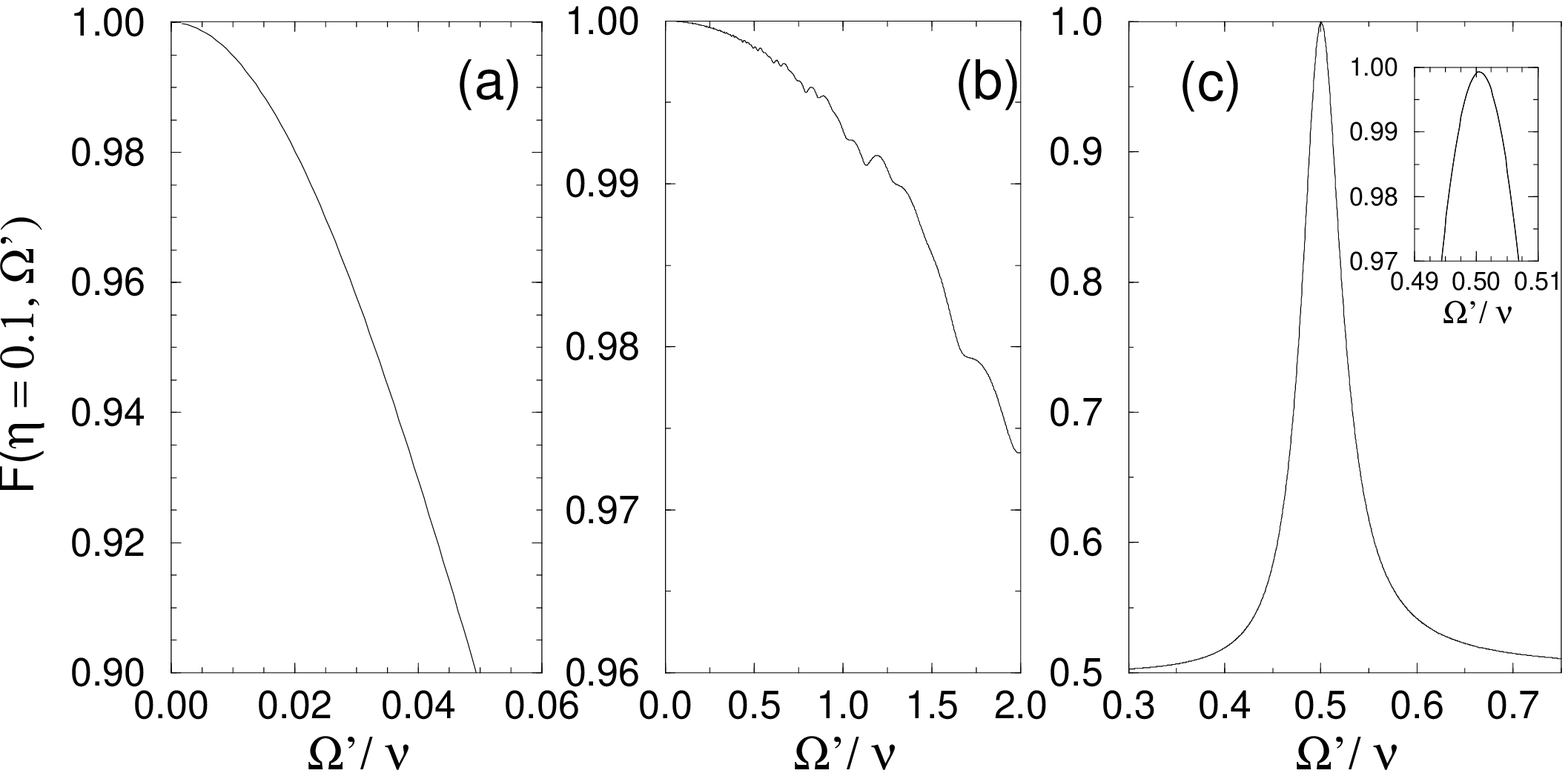}
\caption{\narrowtext Average fidelity measure
 $F(\eta=0.1,\Omega')$ (see section \ref{efficsec}),
plotted against the the ratio $\frac{\Omega'}{\nu}$, for different
quantum gate schemes in a two-ion trap: (a) The travelling-wave CZ
scheme, (b) the standing-wave CZ scheme and (c) the
`lightshift-based' (LB) scheme. Around the resonance
$\Omega'=\frac{\nu}{2}$ (see eq. (\ref{resonance})), the LB scheme
attains a peak efficiency close to $100\%$. The peak stays above
99\% for values of $\frac{\Omega'}{\nu}$ within about 0.5\% of the
resonance (inset). Since $\Omega'$ governs the gate switching
rate, this scheme should allow the implementation of efficient gates over an order
of magnitude faster than those obtained in current experiments based on the
travelling-wave CZ method. Although this is still a few times smaller than the
rate attainable using the standing-wave CZ scheme, the LB
scheme should be easier to implement experimentally.}
\label{effig}
\end{figure}

In Fig. \ref{effig} we plot the efficiency function
F($\eta, \Omega'$), with $\eta$ fixed at 0.1,
 for three different gate schemes: the CZ
scheme using (a)  travelling-wave radiation or (b) standing-wave
radiation; and (c) the `lightshift-based' scheme. The graphs were
obtained by numerical integration of the full Schr\"{o}dinger
equation describing an ion-CM mode interaction in a two-ion trap,
including all off-resonant transitions and all orders of the
Lamb-Dicke parameter. The second or `stretch' mode is assumed to be cooled
to the ground state.

As should be expected, in the CZ schemes the efficiency decreases
essentially monotonically with the laser power. In addition, the
dramatic difference in performance between the standing- and
travelling-wave CZ configurations is readily apparent (note the
difference in scale of the two graphs). Indeed, if we consider
99\% efficiency as the criterion for acceptable gate performance,
then the upper limit for $\Omega'$ in the travelling-wave case is
about $1.5\times 10^{-2}\nu$, while in the standing-wave case
about $1.25\nu$, in agreement with the estimates in eqs.
(\ref{travcond}) and (\ref{standcond}). Meanwhile, the efficiency
of the LB scheme has a narrow peak around the resonance value
$\Omega'=\frac{\nu}{2}$, with a maximum value well over 0.99.
(This is in good agreement with eq. (\ref{epsilon}), which
predicts a population leakage of $\epsilon^2 \sim 0.005$ into the
stretch mode). The width of the region where $F>0.99$ is of the
order of $0.005\nu$. We can conclude that highly efficient gate
performance in this scheme is possible as long as the Rabi
frequency of the laser-ion interaction is stable to within at
least $\pm 0.5\%$.

\subsection{Discussion}

Our results indicate that, as long as the challenges of individual
laser access and ground-state cooling can be met, the
lightshift-based scheme should indeed allow highly efficient
2-qubit gates to be implemented within the Lamb-Dicke regime.
Furthermore, the relatively high
laser power employed in this scheme means these gates should be
over an order of magnitude faster than their counterparts
obtainable via the travelling-wave CZ scheme used in current
experiments. Specifically, an ion-ion C-NOT gate with switching rate
around $\nu/20$ may be realised.
This speed is comparable to the one obtainable in
principle with a standing-wave CZ configuration, but our proposal
achieves it without requiring a precisely controlled standing-wave field. We
believe that these features should make the lightshift-based
scheme a attractive candidate for the realisation of faster
quantum gates. Furthermore, testing the underlying principle of
the scheme in existing single-ion traps should present no
difficulty.

Finally, we would like to briefly compare our scheme with the
`magic Lamb-Dicke parameter' (MLDP) proposal of Monroe {\it et al.}
\cite{Monroe3}. This elegant scheme exploits the fact that the
1-qubit Rabi frequency $\Omega$ in eq. (\ref{1bit}) is in fact
dependent on the number of motional excitations of the ion. It
turns out that, for specific `magic' values of the Lamb-Dicke
parameter, the values of $\Omega$ corresponding to zero and one
phonons become commensurate. This then means that, after a
sufficient number of Rabi periods, the atomic state is flipped or
not depending on the state of the mode, in other words a C-NOT
gate with the mode as control qubit can be implemented. The scheme
has a number of experimental advantages, notably the absence of
the `auxiliary' level needed in the CZ and LB schemes. Also, since
it only uses the strong ionic `carrier' transition, the laser
power used can be quite considerable, leading also to relatively
fast gates. The exact switching rate that can be obtained depends
on the chosen `magic' value, but should be as least as large as
the ones obtained by the other methods discussed in this paper
(see \cite{Steane00} for a discussion).

The method however has also at least two drawbacks. First of all,
even the smallest `magic' value of $\eta$ quoted in \cite{Monroe3}
is 0.316. This is already a bit too large for the validity of the
Lamb-Dicke regime required by currently used cooling mechanisms
such as sideband cooling \cite{Wineland}. Unless more
sophisticated cooling methods are employed (possibly involving the
use of higher-order sidebands \cite{highersideb}), one would then
need the ability to fine-tune $\eta$ to different values at
different stages of the experiment, a feat that has not yet been
accomplished in practice to our knowledge.

A second drawback comes the fact that the MLDP scheme can only
implement universal ion-ion quantum logic \cite{Barenco} if it is
supplemented with another mechanism capable of realising SWAP
gates between internal and motional states. For example, in
\cite{Monroe3}, Monroe et al point out that an ion-ion C-NOT gate
can be realising by ``sandwiching" an MLDP-based ion-mode C-NOT
between two SWAP gates, just as happens in the CZ scheme. However,
the dispersive interaction exploited in the MLDP scheme does not
itself allow the transfer of excitations from the internal to the
motional states. This can be seen by noting that none of the
available gates (1-qubit ionic rotations and C-NOT gates {\em with
the mode as control qubit}) changes the populations in any
motional state. (In other words, these operations alone do {\em
not} constitute a universal set of gates \cite{footnote}.  SWAP
gates can only be realised via some different mechanism, for
instance the CZ red-sideband method or our LB method. In
particular, a gate using LB-based SWAP steps and an MLDP-based
entangling step would combine the best features of both these
schemes, including both speed and the absence of complications
such as auxiliary levels and standing waves (Note though that,
since the LB scheme requires a smaller value of $\eta$, the
ability to tune this parameter would still be required).  Whether
in this ``hybrid" combination or on its own, we hope that the LB
scheme will prove to be a useful tool for ion-trap quantum
information processing.

{\it Endnote}: Shortly after this work was submitted, another
study of the speed limits of Cirac-Zoller gates was put forward by
Steane {\it et al.} \cite{Steane00}. Apart from presenting results
which support and extend the discussion in section IIIA above,
these authors also propose and experimentally test an independent
method for increasing the gate switching rates within a
travelling-wave scenario. Their idea is somewhat complementary to
the one presented in this paper: they argue that the rapid decay
in gate efficiency shown in fig. \ref{effig}a) is partly due to a
shift in the sideband transition frequency caused by the nearby
strong carrier transition. This shift can be compensated for by
choosing the laser beam to be slightly detuned from the first
sideband frequency, resulting in gates that are considerably
faster than the ``standard" CZ gates we have considered in our
analysis. Nevertheless it appears that, if a sufficiently high
gate fidelity is demanded, then our lightshift-based scheme is
still faster than even this enhanced scheme \cite{Steanepriv}.

Acknowledgements: We thank Dana Berkeland, Chris Monroe, H.
Christoph N\"{a}gerl, Juan Poyatos, Danny Segal, Andrew Steane,
J\"{o}rg Steinbach, and Ant\^{o}nio Vidiella-Barranco for
suggestions, clarifications and comments. We acknowledge the
support of the Brazilian agency Conselho Nacional de
Desenvolvimento Cient\'{\i}fico e Tecnol\'{o}gico (CNPQ), the ORS
Award Scheme, the United Kingdom Engineering and Physical Sciences
Research Council, the Leverhulme Trust, the European Science
Foundation, and the European Union.

\appendix

\section{Limits to lightshift gates in N-ion strings}

James \cite{James} has given detailed numerical data for the mode parameters
of up to $10$ trapped ions. It turns out that the frequency $\nu_q$ of a
mode of any given order $q$ is roughly independent of the number of ions (to
about $0.5\%$ over the range of ion numbers investigated). In the second
line of the table below we reproduce these rough frequency values for the
first six modes, relative to the frequency $\nu_1$ of the lowest (CM) mode
\cite{Note2}.

\begin{center}
\begin{tabular}{|c|cccccc|}
\hline
$q$ & 1 & 2 & 3 & 4 & 5 & 6 \\ \hline
$\frac{\nu _{q}}{\nu _{1}}$ & 1 & $\sqrt{3}\simeq 1.73$ & 2.41 & 3.06 & 3.68
& 4.28 \\ \cline{2-7}
$\min \left| _{p}\frac{\left| \nu _{p}-\nu _{q}\right| }{\nu _{q}}\right. $
& $0.73$ & 0.39 & 0.27 & 0.20 & 0.16 &  \\ \cline{2-7}
$\eta_{\max} $ & 0.146 & 0.08 & 0.05 & 0.04 & 0.03 &  \\ \cline{2-7}
$\frac{\eta_{\max}\nu_{q}}{2 \nu_1}$ & 0.073 & 0.069 & 0.065 & 0.061 & 0.055
&  \\ \hline
\end{tabular}
\\[0pt]
\end{center}

We can use this data along with the condition in eq.
\ref{epsilon},i.e.
\begin{equation}
\epsilon^2 = \left(\frac{\eta_{jq}\nu _{q}}{2\left| \nu
_{p}-\nu_{q}\right|}\right)^2 \ll1 \;, \label{cond2}
\end{equation}
in order to estimate the range of values of the Lamb-Dicke parameter $%
\eta_{jq}$ for which the lightshift-based scheme should work
within a given precision. (It can be verified that losses due to
other off-resonant transitions such as the counter-rotating terms
in eq. (\ref{NionHam}) are relatively small in the limit of small
$\eta_{jq}$). For each mode, we list in the third line the
relative frequency spacing to its closest-lying neighbour. Note
that the closest mode is always the next-highest one, and that
their relative spacing {\it decreases} with increasing mode order.
In the fourth line, we list the maximum value $\eta_{\max}$ that
$\eta_{jq}$ can assume such that $\epsilon^2\leq0.01$. Within this
limit we should be able to discard all off-resonant terms in the
Hamiltonian in eq. (\ref{NionHam}), and the dynamics is then well
described by the effective Jaynes-Cummings interaction in eq.
(\ref{effJCM}). Finally, in the fifth line we give the resulting
maximum Rabi frequency achievable using each mode (relative to the
CM mode frequency). Note that the increase of the mode frequencies
themselves is completely compensated by the decrease in the
allowed Lamb-Dicke parameters, with the effect that the overall
Rabi frequency also diminishes as the mode order is increased.

\section{C-NOT gate in the Lightshift scheme}

The following sequence of pulses realises a C-NOT gate between the
internal states of two trapped ions, using the LB ion-mode
interaction given in eqs.(\ref{evol1}-\ref{evol4}).

\begin{enumerate}
\item  First, assuming the `bus' mode is initially in the ground state, the
state of ion 1 {\it in the }$\left| \pm \right\rangle _{1}$ {\it basis } is
mapped onto the $\left| 0\right\rangle $ and $\left| 1\right\rangle $ phonon
states by a 2-qubit $\pi $-pulse of duration $\tau _{1}=\frac{\pi }{\nu
_{q}\eta _{qj}}$%
\begin{mathletters}
\begin{eqnarray}
&&\left| -\right\rangle _{1}\left| 0\right\rangle \stackrel{\tau _{1}}{%
\rightarrow }e^{\frac{i\pi }{2\eta _{jq}}}\left| -\right\rangle _{1}\left|
0\right\rangle \\
&&\left| +\right\rangle _{1}\left| 0\right\rangle \stackrel{\tau _{1}}{%
\rightarrow -}e^{\frac{i\pi }{2\eta _{jq}}}\left| -\right\rangle _{1}\left|
1\right\rangle .
\end{eqnarray}
The phase is identical for both initial states and can be ignored; ion 1 is
left in the $\left| -\right\rangle _{1}$ state. In terms of the logical
basis $\left| g\right\rangle _{1},\left| e\right\rangle _{1}$, this
transformation corresponds to applying a sequence of three gates: first a
Hadamard rotation of the ion, followed by a SWAP\ gate with the mode, and
finally a second Hadamard rotation.

\item  A 1-qubit $\frac{\pi }{2}$ pulse coupling $\left| g\right\rangle _{2}$
to an unpopulated `auxiliary' level $\left| e^{\prime }\right\rangle $ is
then applied on ion 2, mapping $\left| g\right\rangle _{2}\rightarrow \frac{1%
}{\sqrt{2}}\left( \left| g\right\rangle _{2}-\left| e^{\prime }\right\rangle
_{2}\right) \equiv \left| -^{\prime }\right\rangle _{2}$. As in the CZ
scheme, this $\left| g\right\rangle _{2}\leftrightarrow \left| e^{\prime
}\right\rangle _{2}$ transition should be chosen such that level $\left|
e\right\rangle _{2}$ is not affected (for instance by using a different
polarization).

\item  A 2-qubit $2\pi $ pulse of duration $\tau _{2}=\frac{2\pi }{\nu
_{q}\eta _{jq}},$ resonant with the $\left| g\right\rangle
_{2}\leftrightarrow \left| e^{\prime }\right\rangle _{2}$ transition, is
applied on ion 2. States $\left| e\right\rangle _{2}\left| 0\right\rangle
_{2}$ and $\left| e\right\rangle _{2}\left| 1\right\rangle _{2}$ of the
ion-mode system are unaffected by this, while states $\left| -^{\prime
}\right\rangle _{2}\left| 0\right\rangle _{2},\left| -^{\prime
}\right\rangle _{2}\left| 1\right\rangle _{2}$ evolve according to
\end{mathletters}
\begin{mathletters}
\begin{eqnarray}
&&\left| -^{\prime }\right\rangle _{2}\left| 0\right\rangle _{2}\stackrel{%
\tau _{2}}{\rightarrow }\exp \left(
{\displaystyle {i\pi  \over \eta _{jq}}}%
\right) \left| -^{\prime }\right\rangle _{2}\left| 0\right\rangle _{2} \\
&&\left| -^{\prime }\right\rangle _{2}\left| 1\right\rangle _{2}\stackrel{%
\tau _{2}}{\rightarrow }-\exp \left(
{\displaystyle {i\pi  \over \eta _{jq}}}%
\right) \left| -^{\prime }\right\rangle _{2}\left| 1\right\rangle _{2}.
\end{eqnarray}

\item  Another 1-qubit $\frac{\pi }{2}$ pulse coupling $\left|
g\right\rangle _{2}$ to $\left| e^{\prime }\right\rangle _{2}$ is then
applied, mapping $\left| -^{\prime }\right\rangle _{2}$ back to $\left|
g\right\rangle _{2}.$ For convenience, we assume here that this pulse also
cancels the phase acquired in the previous step. The overall effect of the
previous three pulses is to implement a `control-$\sigma _{z}$' gate between
the mode and ion2, which maps
\end{mathletters}
\begin{eqnarray}
\left| g\right\rangle _{2}\left| 0\right\rangle &\rightarrow &\left|
g\right\rangle _{2}\left| 0\right\rangle ;\left| g\right\rangle _{2}\left|
1\right\rangle \rightarrow -\left| g\right\rangle _{2}\left| 1\right\rangle
\\
\left| e\right\rangle _{2}\left| 0\right\rangle &\rightarrow &\left|
e\right\rangle _{2}\left| 0\right\rangle ;\left| e\right\rangle _{2}\left|
1\right\rangle \rightarrow \left| e\right\rangle _{2}\left| 1\right\rangle .
\end{eqnarray}

\item  The state of the mode is then mapped back onto ion 1 by a second $2$%
-qubit $\pi $ pulse
\begin{mathletters}
\begin{eqnarray}
&&\left| -\right\rangle _{1}\left| 0\right\rangle \stackrel{\tau _{1}}{%
\rightarrow }\exp \left(
{\displaystyle {i\pi  \over 2\eta _{jq}}}%
\right) \left| -\right\rangle _{1}\left| 0\right\rangle \\
&&\left| -\right\rangle _{1}\left| 1\right\rangle \stackrel{\tau _{1}}{%
\rightarrow }\exp \left(
{\displaystyle {-i\pi  \over 2\eta _{jq}}}%
\right) \left| +\right\rangle _{1}\left| 0\right\rangle
\end{eqnarray}
\end{mathletters}
\item  Finally, a 1-qubit pulse removes the phase acquired in the previous
step, mapping states $\exp \left(
{\displaystyle {\mp i\pi  \over 2\eta _{jq}}}%
\right) \left| \pm \right\rangle _{1}$ of ion 1 into $\left| \mp
\right\rangle _{1}$. This completes the gate, whose overall effect in the
computational basis is a C-NOT between ion 2 (the control qubit) and ion 1
(the target qubit):
\end{enumerate}
\end{multicols}
\begin{mathletters}
\begin{eqnarray}
&&\left| g\right\rangle _{1}\left| g\right\rangle _{2}\left| 0\right\rangle
\stackrel{1}{\longrightarrow }\left| -\right\rangle _{1}\left|
g\right\rangle _{2}\left( \left| 0\right\rangle -\left| 1\right\rangle
\right) \stackrel{2-4}{\longrightarrow }\left| -\right\rangle _{1}\left|
g\right\rangle _{2}\left( \left| 0\right\rangle +\left| 1\right\rangle
\right) \stackrel{5-6}{\longrightarrow }\left| g\right\rangle _{1}\left|
g\right\rangle _{2}\left| 0\right\rangle ; \\
&&\left| g\right\rangle _{1}\left| e\right\rangle _{2}\left| 0\right\rangle
\stackrel{1}{\longrightarrow }\left| -\right\rangle _{1}\left|
e\right\rangle _{2}\left( \left| 0\right\rangle -\left| 1\right\rangle
\right) \stackrel{2-4}{\longrightarrow }\left| -\right\rangle _{1}\left|
e\right\rangle _{2}\left( \left| 0\right\rangle -\left| 1\right\rangle
\right) \stackrel{5-6}{\longrightarrow }\left| e\right\rangle _{1}\left|
e\right\rangle _{2}\left| 0\right\rangle \\
&&\left| e\right\rangle _{1}\left| g\right\rangle _{2}\left| 0\right\rangle
\stackrel{1}{\longrightarrow }-\left| -\right\rangle _{1}\left|
g\right\rangle _{2}\left( \left| 0\right\rangle +\left| 1\right\rangle
\right) \stackrel{2-4}{\longrightarrow }\left| -\right\rangle _{1}\left|
g\right\rangle _{2}\left( \left| 1\right\rangle -\left| 0\right\rangle
\right) \stackrel{5-6}{\longrightarrow }\left| e\right\rangle _{1}\left|
g\right\rangle _{2}\left| 0\right\rangle ; \\
&&\left| e\right\rangle _{1}\left| e\right\rangle _{2}\left| 0\right\rangle
\stackrel{1}{\longrightarrow }-\left| -\right\rangle _{1}\left|
e\right\rangle _{2}\left( \left| 0\right\rangle +\left| 1\right\rangle
\right) \stackrel{2-4}{\longrightarrow }-\left| -\right\rangle _{1}\left|
e\right\rangle _{2}\left( \left| 0\right\rangle +\left| 1\right\rangle
\right) \stackrel{5-6}{\longrightarrow }\left| g\right\rangle _{1}\left|
e\right\rangle _{2}\left| 0\right\rangle ;
\end{eqnarray}
\end{mathletters}
\begin{multicols}{2}
Note that the first five pulses already generate a `maximally entangling' $2$
-qubit gate, equivalent to the C-NOT gate except for a local rotation.

\end{multicols}

\begin{references}
\bibitem{footnote} Note that excitation transfer would indeed
be possible if 1-qubit rotations could also be implemented
directly on the motional state. This follows since surrounding a
C-NOT gate with Hadamard rotations on both qubits exchanges the
role of control and target qubit, and since a sequence of 3
alternating C-NOT gates is equivalent to the SWAP gate. However,
since the motional state cannot be manipulated independently, a
SWAP gate cannot be realised with this method alone.
\end{references}
\end{document}